\documentclass[RNAAS]{aastex62}
\usepackage{hyperref}
\usepackage{amssymb,amsmath}

\newcommand{\Kepler}{\emph{Kepler}}

\newcommand{\TESS}{\emph{TESS}}

\revised{\today}

\shorttitle{TESS light curves from MIT QLP}
\shortauthors{Huang et al.}

\begin{document}

\title{Photometry of 10 Million Stars from the First Two Years of TESS Full Frame Images}

\correspondingauthor{Chelsea X. Huang}
\email{xuhuang@mit.edu}

\author{Chelsea X. Huang}
\altaffiliation{Torres Postdoctoral Fellow}
\affiliation{Kavli Institute for Astrophysics and Space Research, Massachusetts Institute of Technology, Cambridge, MA, USA 02139}

\author{Andrew Vanderburg}
\affiliation{Department of Astronomy, University of Wisconsin-Madison}

\author{Andras P\'al}
\affiliation{Kavli Institute for Astrophysics and Space Research, Massachusetts Institute of Technology, Cambridge, MA, USA 02139}

\author{Lizhou Sha}
\affiliation{Kavli Institute for Astrophysics and Space Research, Massachusetts Institute of Technology, Cambridge, MA, USA 02139}
\affiliation{Department of Astronomy, University of Wisconsin-Madison}

\author{Liang Yu}
\affiliation{Kavli Institute for Astrophysics and Space Research, Massachusetts Institute of Technology, Cambridge, MA, USA 02139}

\author{Willie Fong}
\affiliation{Kavli Institute for Astrophysics and Space Research, Massachusetts Institute of Technology, Cambridge, MA, USA 02139}

\author{Michael Fausnaugh}
\affiliation{Kavli Institute for Astrophysics and Space Research, Massachusetts Institute of Technology, Cambridge, MA, USA 02139}

\author{Avi Shporer}
\affiliation{Kavli Institute for Astrophysics and Space Research, Massachusetts Institute of Technology, Cambridge, MA, USA 02139}

\author{Natalia Guerrero}
\affiliation{Kavli Institute for Astrophysics and Space Research, Massachusetts Institute of Technology, Cambridge, MA, USA 02139}

\author{Roland Vanderspek}
\affiliation{Kavli Institute for Astrophysics and Space Research, Massachusetts Institute of Technology, Cambridge, MA, USA 02139}

\author{George Ricker}
\affiliation{Kavli Institute for Astrophysics and Space Research, Massachusetts Institute of Technology, Cambridge, MA, USA 02139}

\begin{abstract}
    The Transiting Exoplanet Survey Satellite (TESS) is the first high-precision full-sky photometry survey in space. We present light curves from a magnitude limited set of stars and other stationary luminous objects from the TESS Full Frame Images, as reduced by the MIT Quick Look Pipeline (QLP). 
Our light curves cover the full two-year TESS Primary Mission and include $\sim$ 14,770,000 and $\sim$ 9,600,000 individual light curve segments in the Southern and Northern ecliptic hemispheres, respectively. 
We describe the photometry and detrending techniques we used to create the light curves, and compare the noise properties with theoretical expectations.
All of the QLP light curves are available at MAST as a High Level Science Product via \dataset[10.17909/t9-r086-e880]{\doi{10.17909/t9-r086-e880}}\footnote{\url{https://archive.stsci.edu/hlsp/qlp}}. 
This is the largest collection of TESS photometry available to the public to date.  
\end{abstract}


\section{Introduction}
NASA's Transiting Exoplanet Survey Satellite (TESS) mission launched on 2018 April 18, and began science operations on 2018 July 25. TESS uses four identical cameras to observe stars over a total $24\degr \times96\degr$ field of view. Like its predecessor, \Kepler, TESS observes stars in so-called ``postage stamps,'' or small subimages from the TESS cameras centered on particularly bright or important stars. Unlike \Kepler, however, TESS also saves and downloads images from its entire field of view every 30 minutes. These full frame images (or FFIs) greatly increase the mission's discovery space (with some simulations projecting tens of thousands of planet discoveries from the FFIs \citep{barclay:2018, huang:2018b}, but also substantially increase data processing requirements. 

Before TESS's launch, sophisticated and highly successful software tools were developed to analyze data from space missions like CoRoT \citep{Auvergne:2009} and Kepler \citep{jenkins}. However, because none of these missions downloaded continuous FFIs, the architectures of these pipelines are difficult to adapt to a mission like TESS. In the meantime, unhindered by data downlink rates,  ground-based transit surveys like HAT \citep{Bakos:2007, Bakos:2013}, WASP \citep{Pollacco:2006}, and KELT \citep{Pepper:2007, Pepper:2012} have been extracting light curves and searching for planets from full frame data for many years. These pipelines \citep[e.g.][, etc.]{CollierCameron:2009,Pal:2009} are well suited for analyzing large volumes of wide field images. 

Here, we present light curves for a magnitude limited sample of stars observed by TESS produced by the MIT Quick Look Pipeline (QLP). The QLP, whose architecture includes heritage from both ground-based photometric surveys and methods developed for analysis of space-based photometry, was built to rapidly process TESS data as soon as it is beamed to Earth. The QLP has already been used to identify and alert planet candidates throughout the entire TESS Primary Mission, leading to the discovery of more than 1000 planetary candidates (Guerrero et al. 2020, \textit{submitted to ApJS}), dozens of which have already been confirmed\citep[e.g.][]{huang2018, vanderspek19, Huang:2020, Armstrong:2020}. Here, we describe how QLP produces light curves and evaluate its performance in terms of photometric precision. We publicly release QLP light curves for all sources in the TESS Input Catalog \citep[TIC,][]{stassun2018, stassun2019} observed by TESS in its primary mission down to a limiting TESS magnitude $T$ of 13.5. 

\section{Light Curve Extraction}

\begin{figure*}
    \centering
    \includegraphics[width=0.47\linewidth]{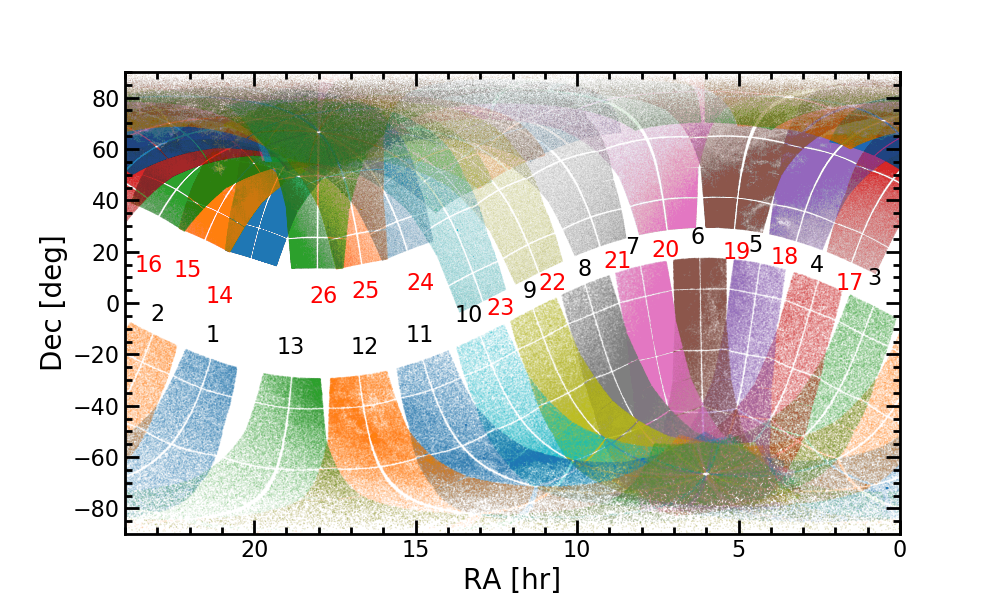}
    \includegraphics[width=0.47\linewidth]{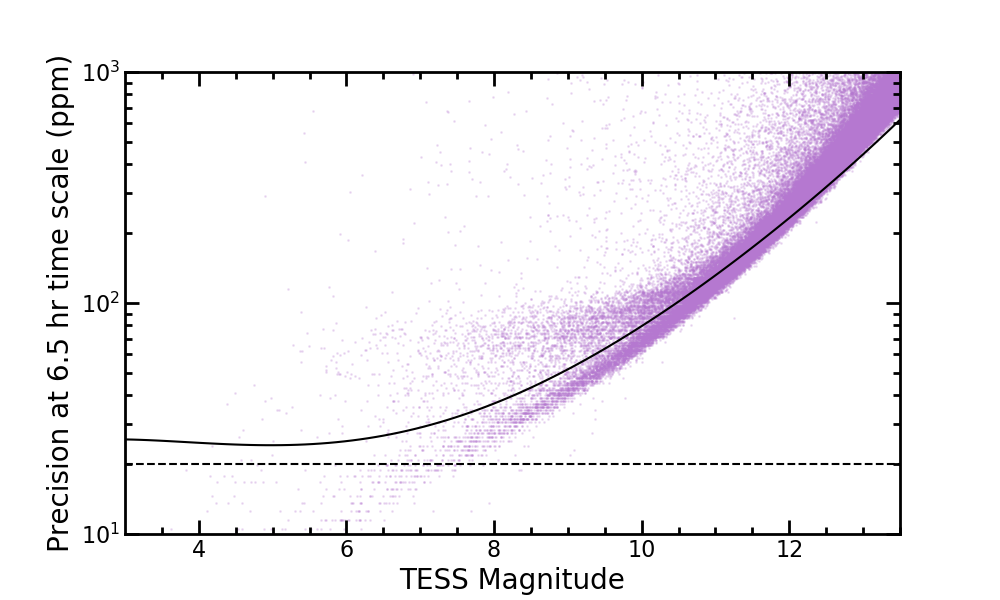}
    \caption{Left: An illustration of all the stars for which we extracted photometry from the TESS Primary Mission FFIs in terms of their equatorial coordinates. Right: The photometric precision of the time series, compared to expected theoretical precision estimated in \citet{sullivan2015} (solid line). The dashed horizontal line is a reference indicating 20 ppm precision. }
    \label{fig:fig1}
\end{figure*}

We extract light curves for all stars in the TIC with TESS magnitude brighter than 13.5 observed in the TESS Primary mission (UT 2018 July 25 - UT 2020 July 04). Figure \ref{fig:fig1} shows the sky locations of stars with QLP light curves. We also added in stars with proper motion larger than 200 $\mathrm{mas}$\,$\mathrm{yr}^{-1}$ and brightness between TESS Magnitude 13.5 and 15.

Our procedure is as follows: 
\begin{enumerate}
    \item We process the TESS FFIs using the TICA software, which will be described by Fausnaugh et~al.\ (in prep.). The TICA software carries out correction of various instrumental effects described in \citet{vanderspek19}. 
    
    \item The images then go through a global background subtraction using nebuliser \citep{Irwin:1985}. Because TESS orbits the Earth, its cameras experience significant levels of scattered background light. Nebuliser removes much of the large-scale scattered background features in the images (features with spatial variation scale $>$20~pixels).
    
    \item We then determine an astrometric solution for each image following \cite{Huang:2015} by relating the measured centroids of bright sources ($8<T<10$) to their coordinates in the TIC. This astrometric solution is then used to obtain the precise position of all catalog sources in the images. The typical astrometric precision for each frame is better than 0.1 pixel ($2\farcs1$).
    
    \item A reference image is constructed with the median combination of 40 good quality images (images with minimal scattered light, as well as good pointing stability), which we identify within each TESS orbit. We then compute difference images for each frame by directly subtracting the reference frame from each individual frames.
    
    \item  We measure the differential source brightness from the difference images by summing the flux within a series of 5 circular apertures with different radii (1.75, 2.5, 3.0, 3.5 and 8.0 pixels) centered on these positions. We continue to process the light curves from all five different apertures identically.

    \item We then perform a second step of background subtraction. Nebuliser effectively removes large background variations, but leaves in some small-scale features. We therefore estimate the background light levels in the difference images at the position of each source by calculating the median pixel values (with iterative outlier rejection) within annuli around the target stars. The background annuli have inner radius and width of 4.0/3.0, 4.0/3.0, 4.0/3.0, 5.0/4.0 and 10.0/5.0 pixels, respectively. We subtract this background estimate from the difference fluxes for each source. 
    
    \item Finally, we convert the measured difference fluxes into absolute fluxes by adding the expected flux from each source based on the TESS-band magnitude estimates from the TIC and the instrument zeropoint magnitude \citep{handbook}. This is equivalent to deblending the flux time series assuming the variations observed are from the target star. 
    
    \item The time series for each light curve is corrected to the Solar System barycentric reference frame from the TESS spacecraft reference frame following the procedure described by \citet{Eastman:2010} Sector 2.1 using the coordinates of the target star at Epoch J2000, and TESS orbits vectors from JPL horizon \footnote{\url{https://ssd.jpl.nasa.gov/horizons.cgi}}.
\end{enumerate}

The time series extracted from this step is called the ``raw light curve" and are post-processed to prepare for the needs of transit search.

\section{Light Curve Post-Processing}

TESS light curves usually contain low-frequency variability from stellar activity or instrumental noise, which must be filtered before the small, short-duration signals caused by transiting planets can be readily detected. We therefore detrend the light curves by applying a high-pass filter before they are searched for transits. Before detrending, we reject outliers in the light curve using the quaternion time series \footnote{The quaternion series can be downloaded from \href{TESS engineering data}{https://archive.stsci.edu/missions/tess/engineering/}. For more details see \citep{Vanderburg:2019}}. Any exposure corresponding to abnormal amplitudes of scatter in the quaternion time series are not used in the detrending. We then fit the light curves from each spacecraft orbit with a basis spline \citep{Vanderburg:2014}. We choose the spacing of the spline break points by minimizing the Bayesian Information Criterion following \citet{shallue}, imposing a minimum allowed spacing of 0.3 days. This minimum spacing is optimized for the detection of planets with short orbital periods; more aggressive splines would lead to the distortion of long-duration transits.

After detrending, we identify anomalous exposures where light curves tend to be noisier than expected based on TESS's typical photometric precision. For each exposure, we look at the set of stars with TESS magnitude between 9.5 and 10 and calculate the fraction which exhibit photometric precision more than 75\% worse than the pre-flight anticipated photometric precision of 200 parts per million (ppm) per hour for stars with $T=10$~mag. If more than 20 \% of the stars observed on the same CCD have poor precision at a particular exposure, we assign a bad quality flag to the corresponding exposure. This bad quality flag is stored as bit 13 in the QUALITY column provided in the light curve products. The rest of the bits are adopted from the SPOC Full Frame Image headers \citep{Jenkins:2015, Jenkins:2016}.    

We combine the detrended light curves observed in two TESS orbits of each TESS sector by offsetting the median of the light curves to the expected TESS magnitude, after rejecting bad quality points. These magnitude time series are then converted to normalized flux time series in the final light curve products.    

The QLP produces light curves of each source from up to five apertures. We identify an optimal light curve for each target based on the source's brightness. Early in the mission, we calculated the photometric precision in each aperture for a set of stars and determined the aperture size that yielded the best photometric precision as a function of TESS-band magnitude (in 13 evenly spaced bins between TESS mag of 6 to 13.5). We used these results to select the optimal aperture. We provide ``raw light curve" from the optimal aperture in the FITS file of QLP HLSP product on MAST under keywords {\bf SAP\_FLUX}. We provided flattened light curves from three apertures with different sizes. The light curve from the optimal aperture is under FITS keyword {\bf KSPSAP\_FLUX}. The light curve from the relatively bigger/smaller apertures are under FITS keyword {\bf KSPSAP\_FLUX\_LAG} and {\bf KSPSAP\_FLUX\_SML}, respectively.

We show the precision of our light curves in Figure \ref{fig:fig1}. For comparison, the solid line is the theoretical photometric precision estimated by \citet{sullivan2015} scaled to a 6.5 hr time scale assuming Gaussian noise. 
The photometric precision roughly follows the predictions for the majority of the stars and has a lower noise floor (approximately 20 ppm) for the brightest stars when the spacecraft operates nominally \footnote{The in-orbit performance of the TESS photometers is better than preflight calculations by \citet{sullivan2015}, which has been traced to an underestimation in their assumed telescope aperture.}.

\section{Caveats} 

QLP light curve production depends critically on the \TESS\ band magnitudes estimated by the TIC. If a star's \TESS\ band magnitude is incorrect, the amplitude of features in the QLP light curve will be likewise incorrect because the QLP will deblend the light curve incorrectly. The uncertainties of the amplitude of variations in the flux time series therefore depend on the uncertainties in the TESS magnitudes. This is not represented in the error-bars we provide in the light curve time series. 
Instead, the uncertainties are estimated with the Median Absolute Median Deviation statistics of each orbit of light curves multiplied by 1.4826.

We note that light curves from Sectors 1--13 were produced using the TIC version 7, while light curves in Sector 14 onward were produced using TIC version 8.  A small fraction of stars have different estimated \TESS\ magnitudes in TIC 7 versus TIC 8. These changes in magnitude affect the amount of deblending applied by QLP and thus the amplitude of light curve features.  

Our method of deblending the light curve time series also do not take into account that the target stars' point spread functions may be not fully contained in the smallest circular aperture (with a radius of 1.75 pixel), leading to underestimation of signal amplitudes in this particular aperture for a large fraction of stars. 

The TICA software we used to calibrate the TESS raw FFIs went through many iterations and version changes during the TESS Primary Mission. We did not keep a record of the particular versions of the calibration software used for each sector of data. We expect this issue to be resolved with future data reprocessing and releases. 
Errors in the TICA smear correction estimate affect the light curves of a small number of stars for one specific version of TICA, so the smear correction was disabled for QLP processing in later sectors.
Users can examine light curves of stars located in the same column to identify such contamination.      

\acknowledgements
We thank the entire TESS Mission team for years of effort to make this work possible. 
This paper includes data collected by the TESS mission, which are publicly available from the Mikulski Archive for Space Telescopes
(MAST). Funding for the TESS mission is provided by NASA's Science Mission directorate.
CXH acknowledges support from MIT's Kavli Institute as a Torres postdoctoral fellow.

\software{%
    nebuliser \citep{Irwin:1985},
    FITSH \citep{fitsh},
    Golang \citep{meyerson2014go},
    Numpy \citep{numpy}, 
    Scipy \citep{2020SciPy-NMeth}, 
    Astropy \citep{astropy:2013, astropy:2018}}

\facility{\TESS{}}
    
\bibliographystyle{aasjournal}

\bibliography{ref}

\begin{figure*}
    \centering
    \includegraphics[width=\linewidth]{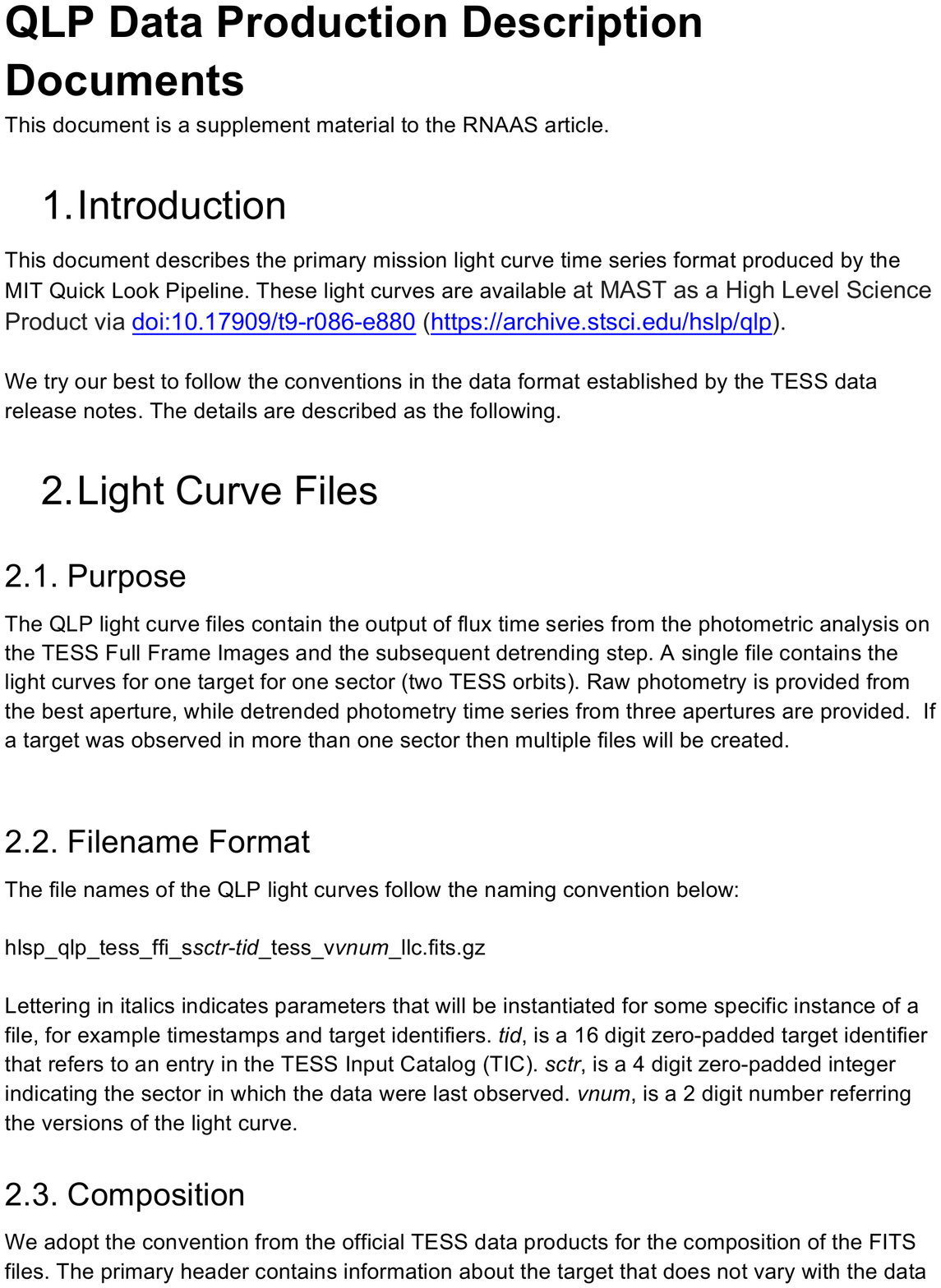}
\end{figure*}
\begin{figure*}
    \centering
    \includegraphics[width=\linewidth]{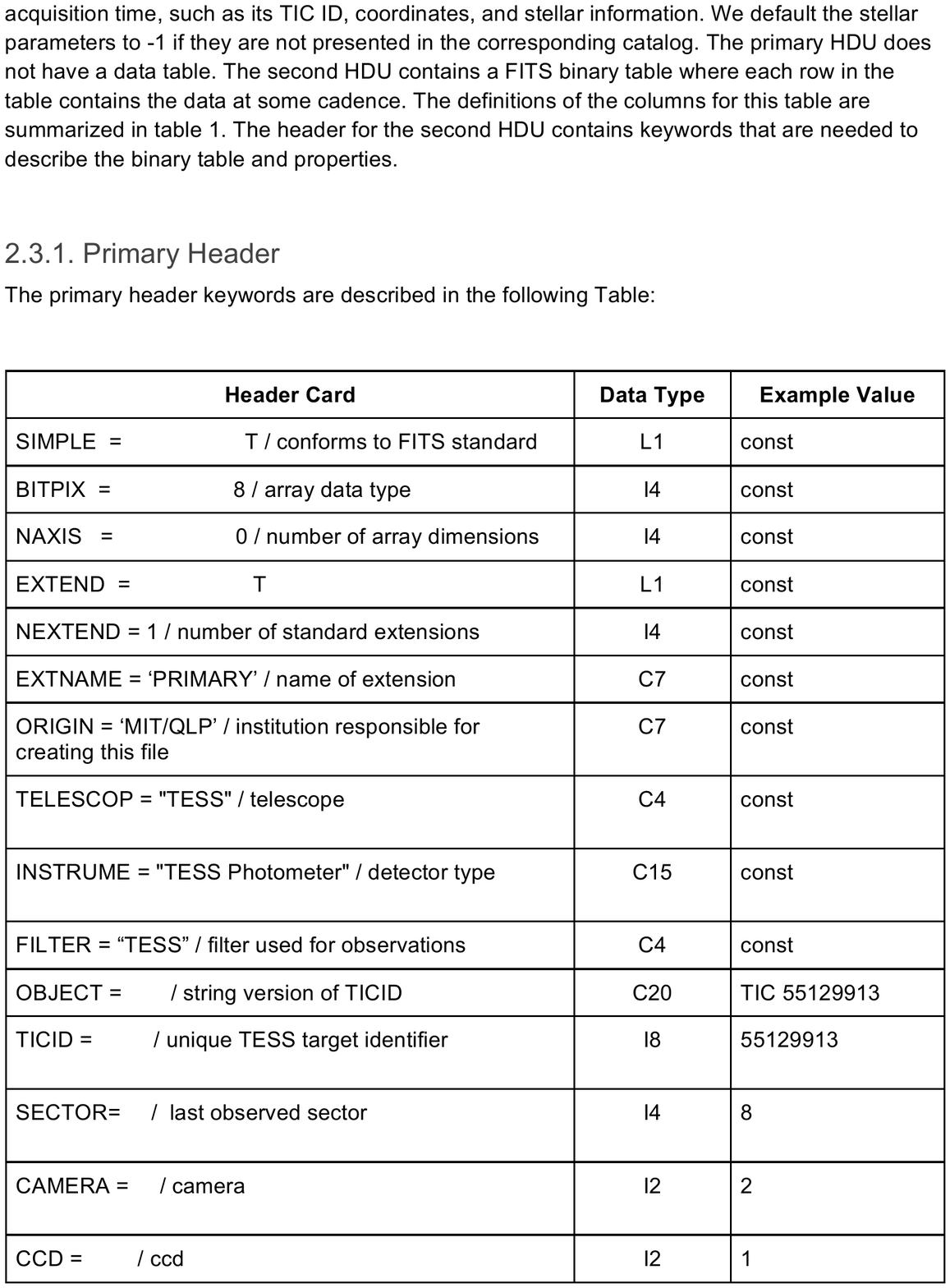}
\end{figure*}
\begin{figure*}
    \centering
    \includegraphics[width=\linewidth]{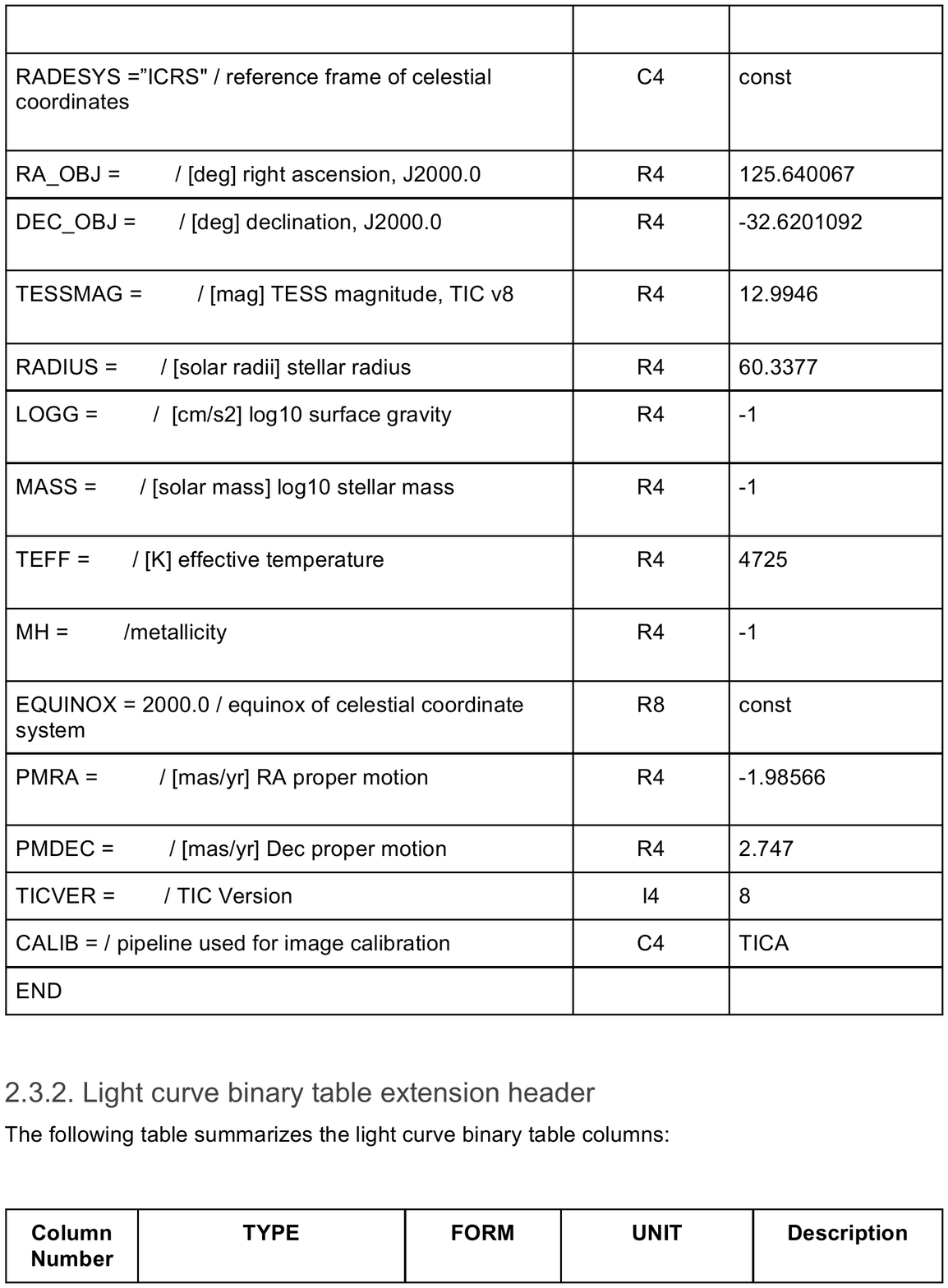}
\end{figure*}
\begin{figure*}
    \centering
    \includegraphics[width=\linewidth]{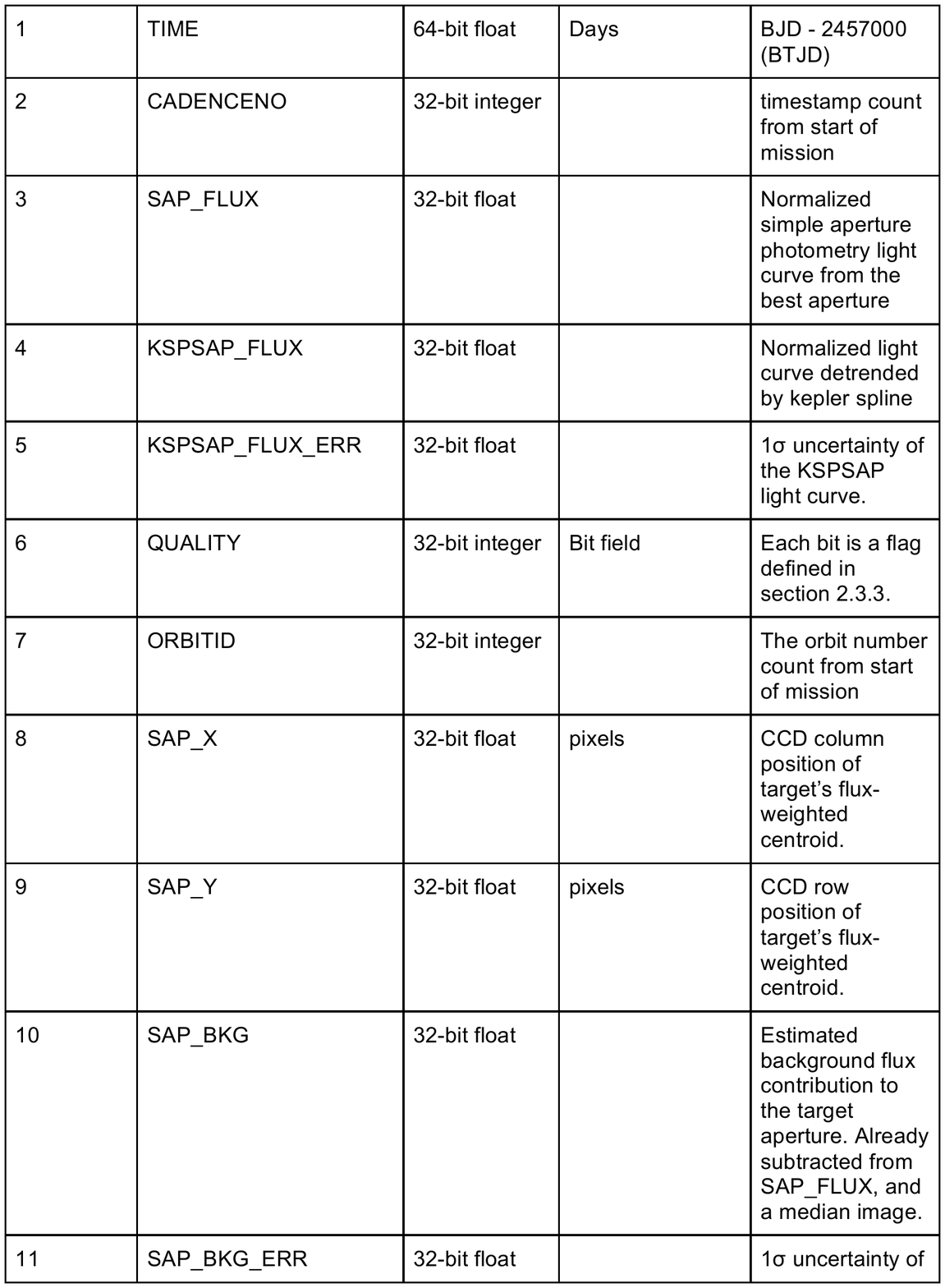}
\end{figure*}
\begin{figure*}
    \centering
    \includegraphics[width=\linewidth]{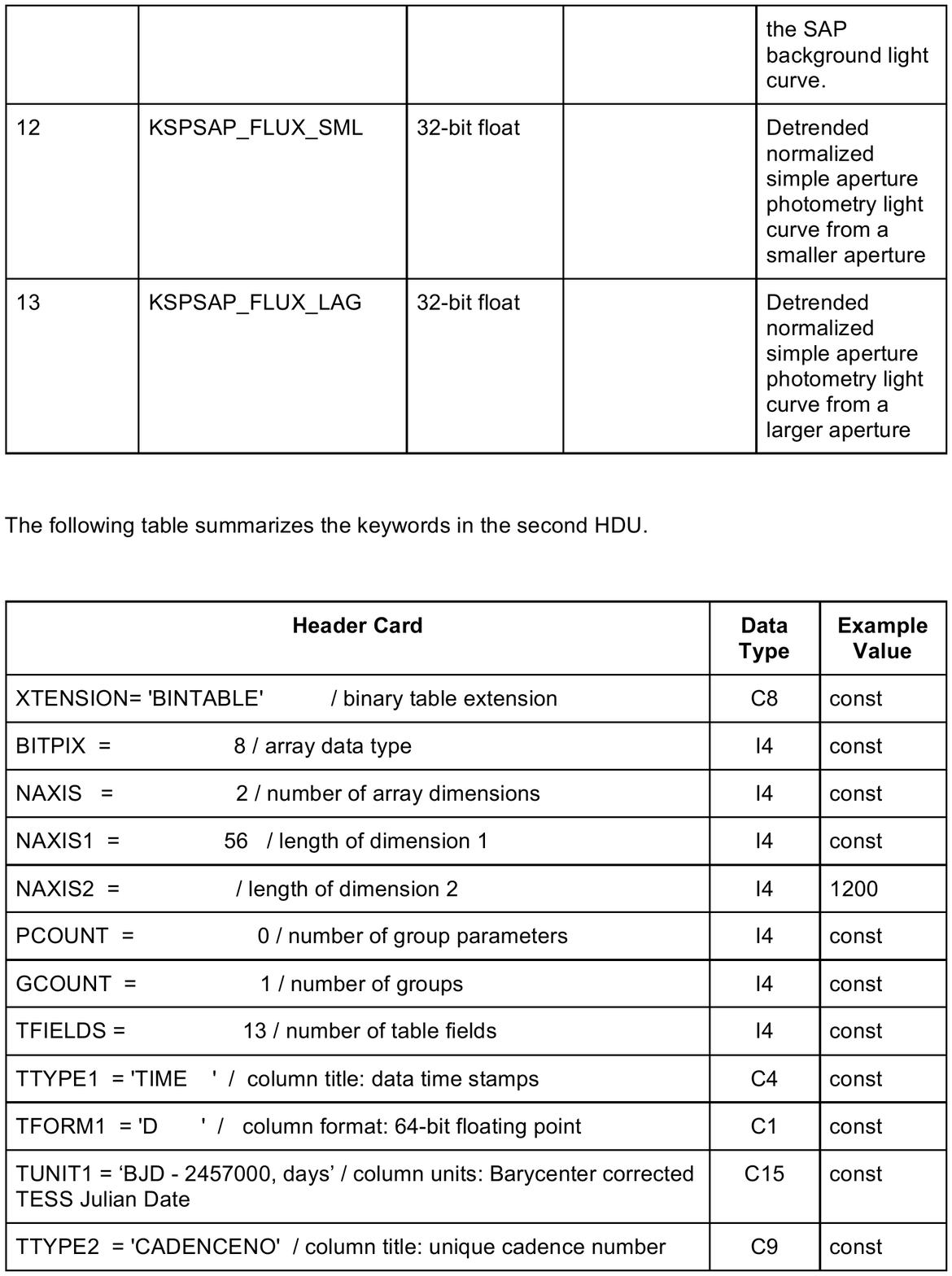}
\end{figure*}
\begin{figure*}
    \centering
    \includegraphics[width=\linewidth]{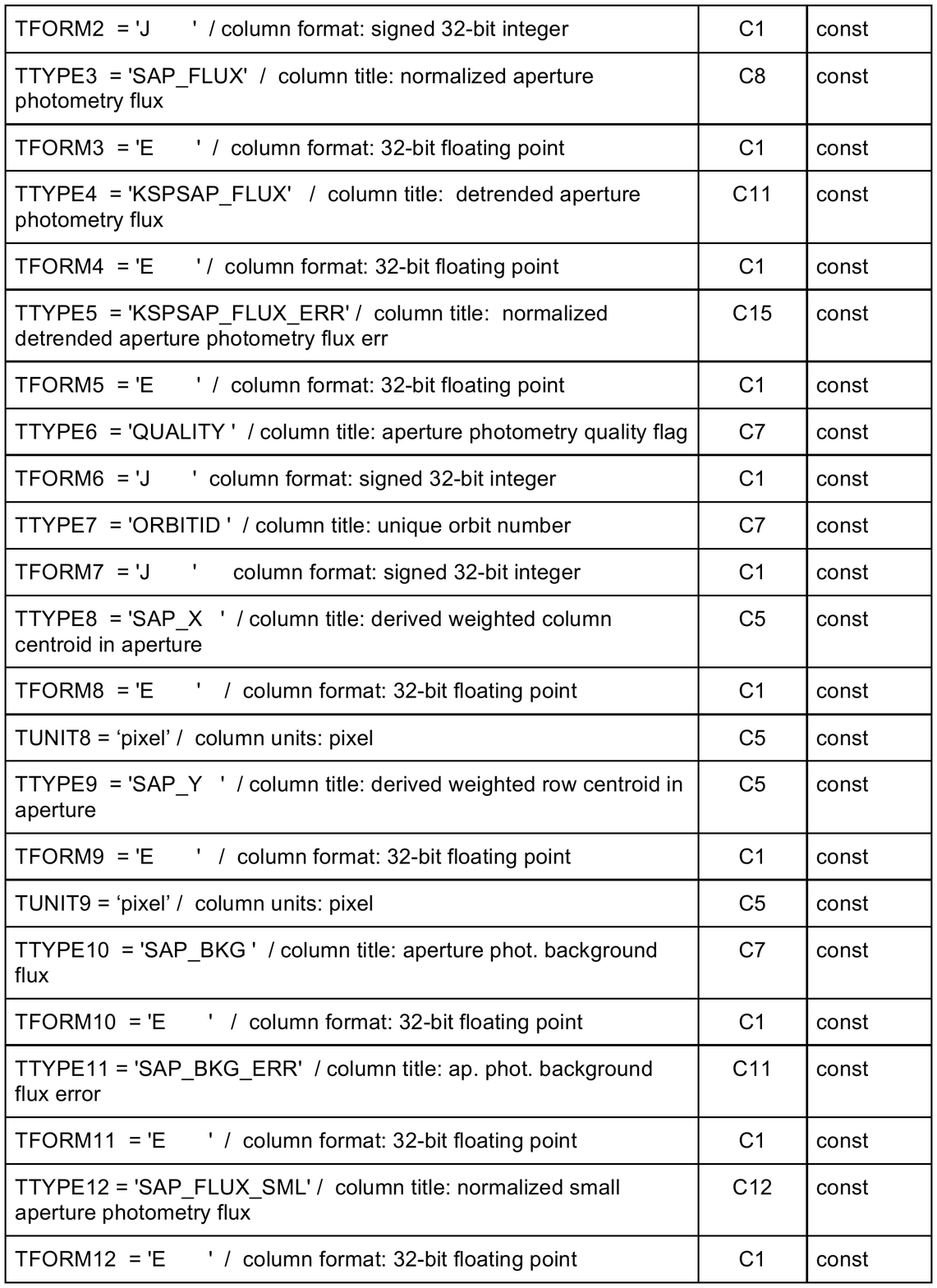}
\end{figure*}
\begin{figure*}
    \centering
    \includegraphics[width=\linewidth]{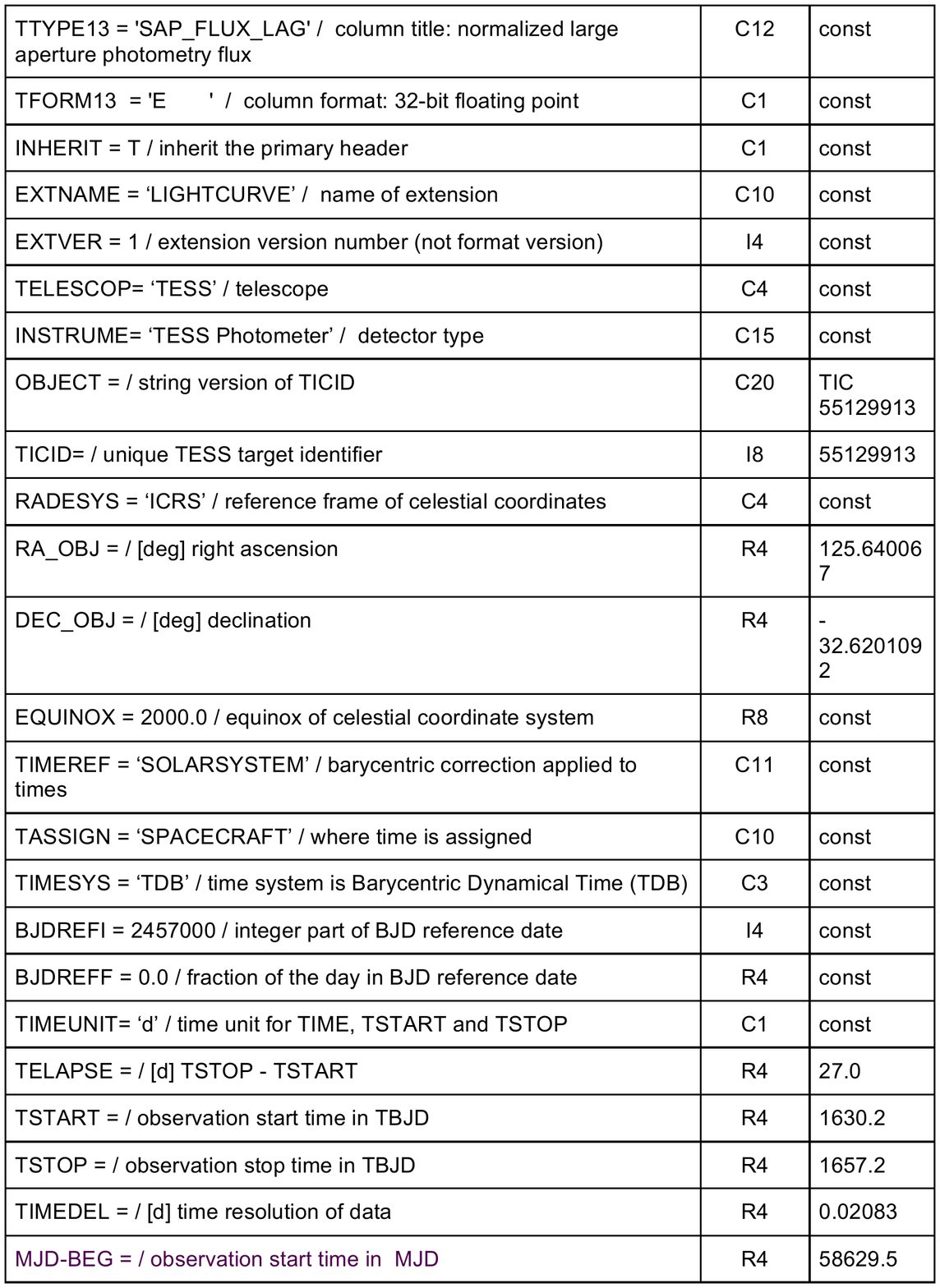}
\end{figure*}
\begin{figure*}
    \centering
    \includegraphics[width=\linewidth]{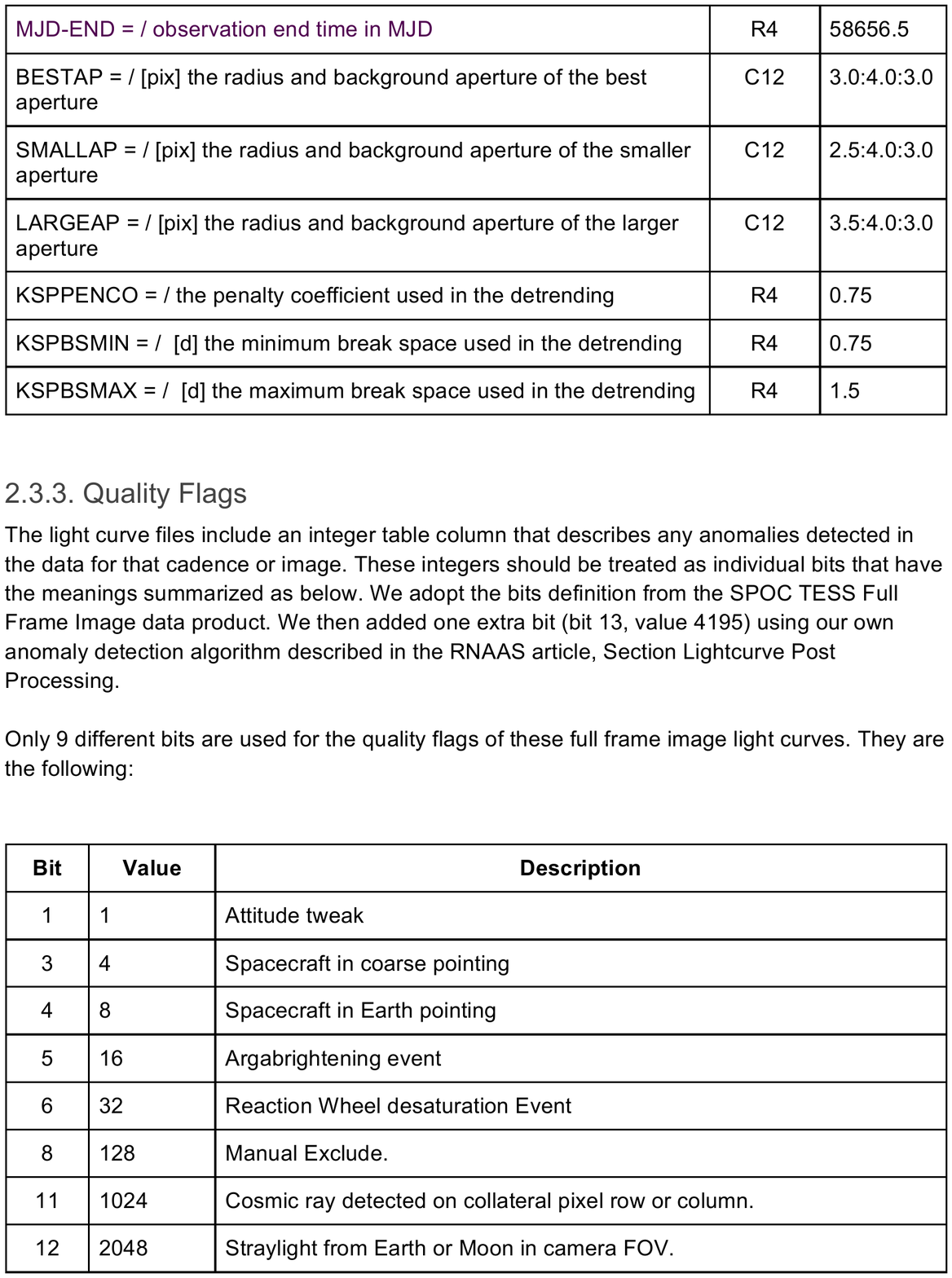}
\end{figure*}
\begin{figure*}
    \centering
    \includegraphics[width=\linewidth]{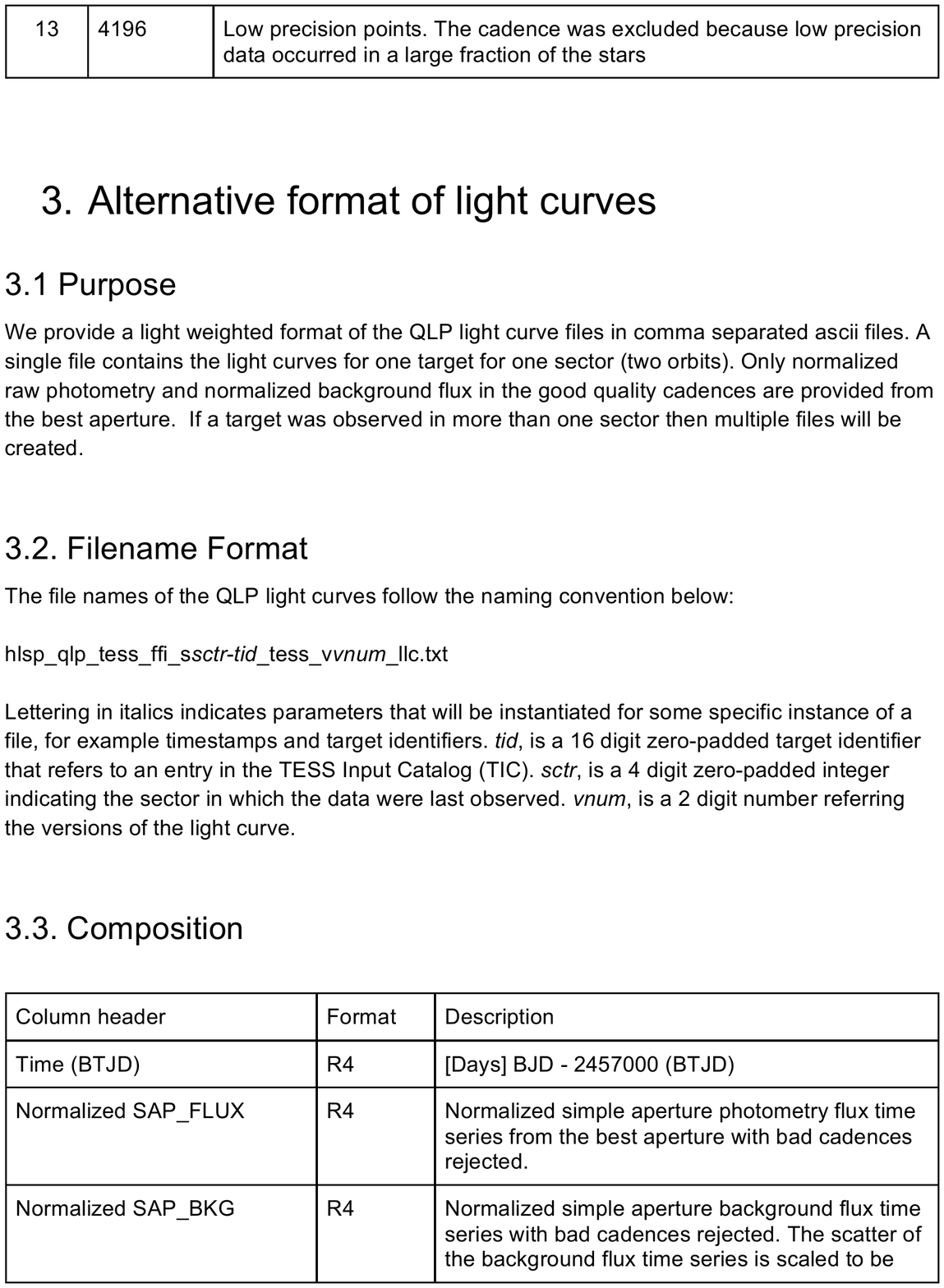}
\end{figure*}
\begin{figure*}
    \centering
    \includegraphics[width=\linewidth]{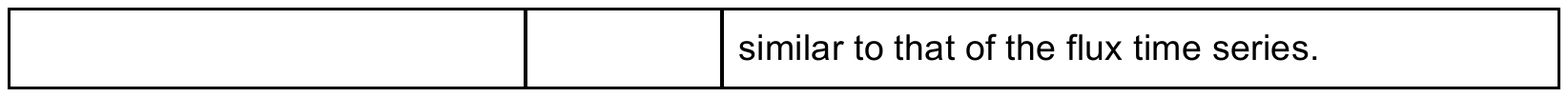}
\end{figure*}

\end{document}